\documentclass[final, runningheads]{llncs}
\usepackage{listings}
\usepackage{color}

\definecolor{mygray}{rgb}{0.5,0.5,0.5}
\definecolor{myblue}{rgb}{0.1,0.1,0.9}
\definecolor{mylightgray}{rgb}{0.95,0.95,0.95}

\usepackage[T1]{fontenc}
\usepackage{graphicx}
\usepackage{amsmath}
\usepackage[backend=biber]{biblatex}
\newcommand{\itadata}{\footnotesize \textsl{ITADATA2024: The 3$^{\text{rd}}$ Italian Conference on Big Data and Data Science}}
\usepackage{fancyhdr}
\fancypagestyle{empty}{\fancyhf{}\fancyhead[C]{\itadata}
  \fancyhead[R]{\includegraphics[width=.05\textwidth]{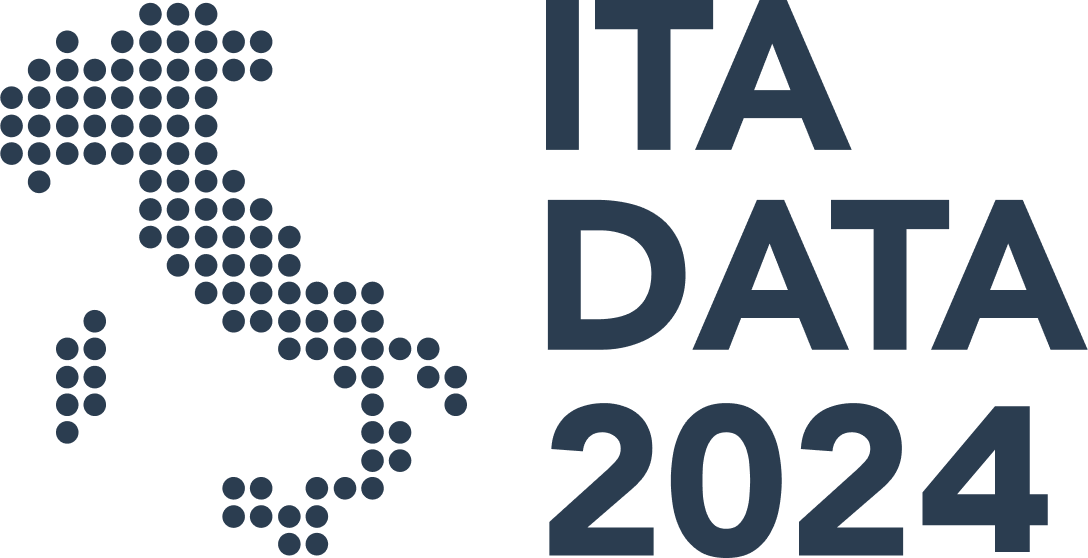}}}

\makeatletter
\begin{document}

\title{Efficient Data Ingestion in Cloud-based architecture: a Data Engineering Design Pattern Proposal}
\author{Chiara Rucco\inst{1}\orcidID{0009-0000-4067-0955} \and
Antonella Longo\inst{1}\orcidID{0000-0002-6902-0160} \and
Motaz Saad\inst{2}\orcidID{0000-0002-1080-7276}}
\authorrunning{F. Author et al.}
\institute{University of Salento, Department of Engineering Innovation
\email{\{chiara.rucco,antonella.longo\}@unisalento.it}
\and
University of Salento, Department of Engineering Innovation\\
\email{motaz.saad@gmail.com}}
\maketitle              \begin{abstract}
In today's fast-paced digital world, data has become a critical asset for enterprises across various industries. However, the exponential growth of data presents significant challenges in managing and utilizing the vast amounts of information collected. Data engineering has emerged as a vital discipline addressing these challenges by providing robust platforms for effective data management, processing, and utilization.
Data Engineering Patterns (DEP) refer to standardized practices and procedures in data engineering, such as ETL (extract, transform, load) processes, data pipelining, and data streaming management. Data Engineering Design Patterns (DEDP) are best practice solutions to common problems in data engineering, involving established, tested, and optimized approaches. These include architectural decisions, data modeling techniques, and data storage and retrieval strategies. While many researchers and practitioners have identified various DEPs and proposed DEDPs, such as data mesh and lambda architecture, the challenge of high-volume data ingestion remains inadequately addressed.
In this paper, we propose a data ingestion design pattern for big data in cloud architecture, incorporating both incremental and full refresh techniques. Our approach leverages a flexible, metadata-driven framework to enhance feasibility and flexibility. This allows for easy changes to the ingestion type, schema modifications, table additions, and the integration of new data sources, all with minimal effort from data engineers. Tested on the Azure cloud architecture, our experiments demonstrate that the proposed techniques significantly reduce data ingestion time.
Overall, this paper advances data management practices by presenting a detailed exploration of data ingestion challenges and defining a proposal for an effective design patterns for cloud-based architectures.

\keywords{Data Ingestion  \and Design Patterns \and Cloud}
\end{abstract}

\section{Introduction}

Data engineering involves designing and building systems for collecting, storing, and analyzing data at scale, ensuring that data is accessible, reliable, and ready for analysis \cite{jain}. In data pipelines, design patterns are crucial for creating scalable and robust data architectures, capable of handling large volumes of data generated by modern enterprises \cite{sabatucci}. By adhering to design pattern best practices, organizations can enhance their data pipeline architecture, saving both time and money. The successful application of design patterns in areas like Service-Oriented Architecture (SOA) underscores their value in achieving high-quality designs across infrastructures \cite{wedyan}.

Data Engineering Patterns (DEPs) refer to standardized practices in data engineering, including ETL (extract, transform, load) processes, data pipelining, and data streaming management \cite{DEDPBook2024}. Data Engineering Design Patterns (DEDPs) are best practice solutions to common problems in data engineering, involving established, tested, and optimized approaches \cite{DEDPBook2024}. These patterns address architectural decisions, data modeling techniques, and data storage and retrieval strategies \cite{chen}. DEDPs offer business users and technical experts methods to tackle recurring issues with repeatability, consistency, and structure.

The exponential growth of data presents significant challenges, particularly in data ingestion—a critical phase of the data pipeline. Efficiently ingesting high-volume data from diverse sources such as databases, files, and APIs is complex and time-consuming. This challenge is compounded by the need to handle real-time data streams and batch processes, requiring solutions that adapt to varying data types and ingestion frequencies without extensive reengineering. Data ingestion is a recurrent challenge among data engineers, demanding innovative approaches that offer flexibility and efficiency \cite{meehan}. While many researchers and practitioners have identified various DEPs and DEDPs such as data mesh and lambda architecture, efficiently ingesting high-volume data remains inadequately addressed \cite{machado}, \cite{munshi}. Existing design patterns often fall short in providing the necessary adaptability and ease of integration for new data sources, schema changes, and different ingestion techniques \cite{sangat}.

This paper provides a comprehensive review of existing efforts on DEDPs, identifying key themes, methodologies, findings, and gaps in research on data ingestion design patterns in cloud architectures. It proposes a pattern to address the limitations of existing approaches in Azure, AWS, and Google Cloud. By offering a robust solution to one of the most persistent challenges in data engineering, this paper highlights the gaps in current literature and demonstrates how the proposed work aims to address these gaps, offering an innovative approach to overcome existing challenges.

The proposed design pattern for data ingestion is tailored for big data in cloud architecture, incorporating both incremental and full refresh techniques within a flexible, metadata-driven framework. This approach enhances the feasibility and flexibility of data ingestion processes, allowing seamless changes to ingestion types, schema modifications, table additions, and the integration of new data sources with minimal effort from data engineers. By leveraging metadata, this design pattern ensures that changes can be managed dynamically without extensive manual intervention, reducing overall data ingestion time \cite{bellini}.

Tested on Azure, the proposed design pattern demonstrates its effectiveness in significantly reducing data ingestion time, validating the approach and highlighting its potential for application across other cloud platforms like AWS and Google Cloud \cite{grover}. By providing a comprehensive framework for data ingestion, this pattern equips organizations with tools to manage their data more effectively, enhance data quality and consistency, and ultimately drive informed decision-making and business growth.

The paper is structured as follows: the second section reviews the state-of-the-art data engineering patterns in cloud-based architectures and data engineering design patterns. The third section delves into various data ingestion methods and existing techniques. The final section presents the proposed design pattern and framework, validated through testing in an Azure cloud-based architecture.

\section{Background and state of art}

In this section, we will review related works on data engineering design patterns, cloud-based data engineering design patterns, and data ingestion in cloud architectures. First, we will review significant studies and contributions in the field of data engineering design patterns, providing a foundation for understanding current trends and methodologies. Following this, we will delve into cloud-based data engineering design patterns, examining how these patterns enable the development of scalable, resilient, and efficient data systems within cloud environments. Lastly, we will explore data ingestion in cloud architectures, focusing on the methodologies and best practices that ensure effective and seamless data flow from various sources into cloud-based data systems.

\subsection{Related works on data engineering design patterns}

Data engineering design patterns are essential for efficiently processing large and diverse datasets. Research has highlighted the importance of well-structured data ingestion pipelines to manage the complexities of handling massive data volumes from various sources. Key basic patterns in this domain include ETL (Extract, Transform, Load) and ELT (Extract, Load, Transform). More advanced patterns include Lambda Architecture, Kappa Architecture, Data Mesh, and Data Vault \cite{lambda,DataKappa,Datamesh,dataVault}.

The Lambda and Kappa Architectures are proposed to handle both batch and real-time data processing. These architectures provide flexibility and efficiency in managing large datasets \cite{lambda,DataKappa}. The Kappa Architecture is a robust framework for high-speed data processing in various domains like Agriculture 4.0 \cite{DataKappa}, ensemble learning from data streams with concept drift [2], and personal analytics in the Internet of Things (IoT) \cite{DataKappaIoT}. It excels in real-time decision-making by efficiently handling large volumes of data from sensors, machines, and IoT devices. On the other hand, the Data Mesh concept represents a shift towards decentralized data management, promoting autonomy and domain-oriented data ownership \cite{Datamesh}. 

Data mesh architecture emphasizes key principles such as treating data as a product, domain ownership of data, self-serve data platforms, and federated computational governance, which enhance operational efficiency and agility within enterprises \cite{Datamesh}. Additionally, the infusion of AI capabilities in data mesh solutions automates tasks, provides self-service capabilities, and ensures the delivery of trustworthy AI, thereby adding significant value for business users \cite{DataMeshAI}. 

Finally, Data Vault methodologies offer scalable and auditable ways to manage historical data \cite{dataVault}. The Data Vault 2.0 standard, derived from nature's simplicity, helps prevent common data warehousing failures and is successfully applied across organizations of varying sizes. In the modern database landscape, the emergence of NoSQL stores like MongoDB addresses agility and scalability challenges, prompting researchers to develop a metamodel for translating MongoDB into a Data Vault-based enterprise data warehouse, facilitating the integration of diverse data sources and enabling comprehensive tracking of schema changes \cite{NoSQLDataVault}. 

In summary, the adoption of these data engineering design patterns allows organizations to efficiently manage and process large datasets, addressing the challenges posed by data volume and diversity. Table \ref{tab:common_DEDP} shows common data engineering design patterns proposed by researchers and used by practitioners.

\begin{table}[ht]
\centering
\renewcommand{\arraystretch}{1.5}
\label{tab:common_DEDP}
\begin{tabular}{|p{0.15\textwidth}|p{0.4\textwidth}|p{0.45\textwidth}|}
\hline
\textbf{Design Pattern} &
  \textbf{Description} &
  \textbf{When to Use?} \\ \hline
ETL (Extract, Transform, Load) &
  Extracts data from sources, transforms it, and loads it into a target system. &
  Use when transformations need to be done before loading data into the target system. \\ \hline
ELT (Extract, Load, Transform) &
  Extracts data, loads it into a target system, and then transforms it. &
  Use when the target system has strong processing capabilities to handle transformations. \\ \hline
Lambda Architecture \cite{lambda} &
  Combines batch and real-time processing for comprehensive data. &
  Use for systems requiring both real-time updates and historical data processing. \\ \hline
Kappa Architecture \cite{DataKappa} &
  Processes all data in real-time, avoiding a separate batch layer. &
  Use when real-time data processing is a primary requirement and batch processing is less critical. \\ \hline
Data Mesh \cite{Datamesh} &
  Decentralizes data ownership and treats data as a product. &
  Use in large organizations with multiple domains requiring autonomous data management and governance. \\ \hline
Data Vault \cite{dataVault} &
  Models data for flexibility and historical tracking. &
  Use when historical tracking and auditability of data changes are critical. \\ \hline
 
\end{tabular}
 \caption{Common Data Engineering Design Patterns}
\end{table}

\subsection{Cloud based data engineering design patterns}

On the other hand, major cloud providers such as Microsoft Azure, Amazon Web Services (AWS), and Google Cloud Platform (GCP) have developed comprehensive data engineering design patterns tailored to their platforms. These patterns enable businesses to build scalable, secure, and efficient data pipelines. \cite{microsoft_cloud_design_patterns,aws_cloud_design_patterns,gcp_cloud_design_patterns}. 

Microsoft Azure offers patterns like the Data Lake pattern, which leverages Azure Data Lake Storage for managing large volumes of structured and unstructured data. According to \cite{microsoft_cloud_design_patterns}, Azure addresses challenges in cloud design patterns including data management, design and implementation, and messaging. The article lists a catalog of patterns and the categories they belong to, with data engineering-related categories including Cache-Aside, Command and Query Responsibility Segregation (CQRS), Event Sourcing, Index Table, Materialized View, Sharding, and Valet Key as shown in Table \ref{tab:ms_azure_patterns}. Azure offers tools and services like the Data Lake pattern, which leverages Azure Data Lake Storage for managing large volumes of structured and unstructured data, and the Lambda Architecture pattern using Azure Stream Analytics for real-time and batch processing.

\begin{table}[ht]
\centering
\renewcommand{\arraystretch}{1.5}
\begin{tabular}{|p{0.35\textwidth}|p{0.65\textwidth}|}
\hline
\textbf{Pattern} & \textbf{Summary}  \\
\hline
Cache-Aside & Load data on demand into a cache from a data store. \\
\hline
Command and Query Responsibility Segregation (CQRS) & Segregate operations that read data from operations that update data by using separate interfaces.  \\
\hline
Event Sourcing & Use an append-only store to record the full series of events that describe actions taken on data in a domain.  \\
\hline
Index Table & Create indexes over the fields in data stores that are frequently referenced by queries.  \\
\hline
Materialized View & Generate prepopulated views over the data in one or more data stores when the data isn't ideally formatted for required query operations.  \\
\hline
Sharding & Divide a data store into a set of horizontal partitions or shards.  \\
\hline
Valet Key & Use a token or key that provides clients with restricted direct access to a specific resource or service.  \\
\hline
\end{tabular}
\caption{Microsoft Azure Data Management on the Cloud Design Patterns \cite{microsoft_cloud_design_patterns}}
\label{tab:ms_azure_patterns}
\end{table}

AWS offers data-driven architectural patterns derived from commonly seen use cases on AWS. In \cite{aws_cloud_design_patterns}, AWS outlines the five most commonly seen architecture patterns that cover several use cases across various industries and customer sizes: Customer 360 architecture, Event-driven architecture with IoT data, Personalized architecture recommendations, Near real-time customer engagement, and Data anomaly and fraud detection. AWS provides tools and services that utilize these data engineering design patterns (DEDPS), such as the Serverless Data Lake, which uses Amazon S3 for storage, AWS Glue for ETL, and Amazon Athena for querying, as well as the Kappa Architecture with Amazon Kinesis for real-time data processing \cite{aws_cloud_design_patterns}.

Google Cloud Platform (GCP) offers business use cases, sample code, and technical reference guides for industry data analytics, which identify best practices to accelerate the implementation of workloads \cite{gcp_cloud_design_patterns}. GCP provides tools and services that apply DEDPs, such as the Dataflow-based ETL pipeline, which uses Google Cloud Dataflow for stream and batch processing, and the BigQuery Lambda Architecture for integrating batch and streaming data with Google BigQuery \cite{gcp_cloud_design_patterns}.

Although researchers and practitioners have identified various Data Engineering Patterns (DEPs) and proposed Data Engineering Design Patterns (DEDs), such as data mesh and lambda architecture, and despite significant advancements by cloud providers in developing robust data engineering design patterns, a notable gap persists: the challenge of high-volume data ingestion remains inadequately addressed. There is substantial room for improvement in handling data ingestion from diverse sources requiring different processing techniques.

In this paper, we propose a data ingestion design pattern for big data within cloud architecture, incorporating both incremental and full refresh techniques. Our approach leverages a flexible, metadata-driven framework to enhance feasibility and adaptability. This allows for easy modifications to the ingestion type, schema updates, table additions, and integration of new data sources, all with minimal effort from data engineers.

\subsection{Data Ingestion in Cloud Architectures}

In a cloud-based architecture, data ingestion is pivotal for collecting, importing, and integrating data from diverse sources into centralized repositories like data lakehouses, ensuring accessibility and reliability for subsequent analysis. Managing the exponential growth in data volume and velocity demands robust techniques for efficient processing.
Organizations must handle large volumes of data and process it in real-time to keep up with the rapid pace of data generation. Techniques such as batch processing and real-time streaming are employed to capture and process data streams efficiently. 

\textbf{Traditional Batch Processing} remains fundamental for handling large-scale data ingestion tasks that do not require immediate availability. Batch processing is particularly suited for historical data analysis, data warehousing, and scheduled updates. It allows organizations to process vast amounts of data in bulk, ensuring efficiency and structured management of data flows \cite{sangat}.

\textbf{Real-Time Streaming} technologies have become indispensable for scenarios requiring instant data processing and low-latency applications. Tools such as Apache Kafka, Apache Flume, and AWS Kinesis Data Streams enable continuous data ingestion and processing, supporting applications like fraud detection, real-time analytics, and IoT data processing \cite{grover}. These technologies ensure that data is available for immediate insights and operational decision-making, enhancing agility and responsiveness in dynamic environments.

Another challenge is the variety and heterogeneity of data. Data comes in various formats, types, and structures, including structured, unstructured, and semi-structured data. To handle this diversity, robust mechanisms are needed to extract, transform, and load data from different sources while ensuring data integrity. Ensuring data quality and reliability is also critical during the data ingestion process. Data cleansing, validation, and error handling techniques are employed to address issues related to data completeness, accuracy, and consistency. By maintaining high data quality, organizations can rely on the ingested data to generate accurate insights and make informed decisions. \cite{ingestion}

\subsection{Tools and techniques used for Data Ingestion}
Data ingestion is a critical component of cloud-based architectures, enabling the transfer, processing, and integration of data from various sources into centralized systems for analysis and storage. Here we delve into the state-of-the-art techniques in data ingestion, focusing on full ingestion and incremental ingestion methodologies, and elaborates on the advantages and challenges of each approach.

\subsubsection{Full Ingestion}
In cloud-based architectures, full ingestion involves transferring entire datasets from source systems to the destination cloud environment. This method is particularly useful during initial data loads or periodic full data refreshes to ensure data consistency across the platform. One of the primary advantages of full ingestion is that it ensures data consistency by synchronizing the entire dataset between the source and destination, thereby reducing the risk of discrepancies. Additionally, full ingestion simplifies data management processes, as there is no need to track individual changes within the dataset. This approach, however, is not without its challenges. Full ingestion can be resource-intensive, requiring significant computational power, storage, and network bandwidth, especially when dealing with large datasets. The process can also be time-consuming, leading to potential downtime and delays in data availability.  Meehan et al. \cite{meehan} discuss a three-layer (edge-fog-cloud) architecture for data management in IoT applications, highlighting the use of full ingestion for initial data loads and emphasizing the importance of data consistency and reliability in such environments.

\subsubsection{Incremental Ingestion}
On the other hand, incremental ingestion focuses on transferring only the data that has changed since the last ingestion cycle. This method leverages techniques like change data capture (CDC) to identify and process modified records, significantly reducing the load on the system. Incremental ingestion has several advantages, including reduced system load and the ability to provide near real-time updates, making it ideal for applications that require timely information. By only transferring changed data, incremental ingestion minimizes the computational and network resources required. However, implementing incremental ingestion can be complex, requiring sophisticated mechanisms to track and capture data changes accurately. Additionally, if not managed correctly, there is a risk of data inconsistencies between the source and destination. Marcu \cite{marcu} explores the use of incremental ingestion in a scalable architecture designed for high throughput and low latency, highlighting its benefits in big data environments where timely data processing is crucial.

\subsubsection{Hybrid Models}
Modern cloud architectures often implement a hybrid model, combining both full and incremental ingestion techniques to balance the need for up-to-date information with system performance and resource optimization. Hybrid models leverage the strengths of both full and incremental ingestion, optimizing resource utilization and performance. This approach provides flexibility, allowing organizations to tailor their data ingestion strategy to their specific needs. For instance, full ingestion can be used for initial loads, while incremental ingestion can handle ongoing updates. Marcu et al. \cite{marcu2018survey} discuss a unified ingestion and storage architecture that supports both full and incremental ingestion methods, arguing that this hybrid approach is essential for efficiently managing data in cloud-based systems.

\section{Data Ingestion Design Pattern proposal}

In this section, we propose a cloud-agnostic design pattern for a robust and scalable data ingestion framework, incorporating best practices to handle diverse data sources. This design pattern aims to address the challenges associated with ingesting data from various sources such as Oracle databases, Excel files, SQL databases, APIs, and more. By employing this pattern, organizations can streamline their data ingestion processes into a Data Lake, which serves as the foundational storage layer in a Data Lakehouse architecture.

The key objective of this design pattern is to combine the benefits of both incremental and full refresh data ingestion techniques. Recognizing that different data sources have varying capabilities and update frequencies, our approach leverages the most efficient ingestion method for each source. This flexibility ensures optimal performance and resource utilization, adapting to the unique characteristics of each data source.

A crucial component of this design pattern is the use of a metadata-driven approach to manage and orchestrate the data ingestion process. We propose the implementation of a mapping table, stored in a SQL database, which contains essential metadata for each table to be ingested as shown in Table \ref{tab:example-table}. This metadata includes the data source name, table name, configuration parameters (such as credentials), primary key information (if available), date column for incremental ingestion (if available), and the desired type of ingestion (full or incremental).

\begin{table}[ht]
\centering
\renewcommand{\arraystretch}{1.5} 
\begin{tabular}{|c|c|c|c|c|c|}
\hline
Data Source Name & Table Name & Credentials & Primary Key & Watermark & Ingestion Type \\
\hline
DataSource1 & Table1 & Credentials1 & PK1 & WM1 & Full  \\
DataSource2 & Table1 & Credentials2 & PK2 & WM2 & Incremental \\
DataSource2 & Table2 & Credentials3 & PK3 &  & Incremental  \\
\hline
\end{tabular}
\caption{Mapping Table}
\label{tab:example-table}
\end{table}

This metadata-driven approach provides a high level of customization and control over the data ingestion process, allowing organizations to define the ingestion method based on the capabilities and characteristics of each data source. By utilizing a metadata table, we can dynamically adjust the ingestion strategy, ensuring a balance between efficiency and comprehensiveness.
The design pattern also emphasizes a systematic approach to extract, transform, and load (ETL) data from each source into the Data Lake. By adopting a folder-based approach within the Data Lake, each table is stored in a separate folder, and the use of formats with transactional capabilities and efficient data storage and retrieval, such as Delta Lake, is recommended.

To handle the actual data ingestion and processing, this design pattern proposes the use of techniques based on a date column and ingestion based on a hash. These techniques provide flexibility and efficiency in managing data ingestion scenarios:

\textbf{Date Column-Based Ingestion}: Leveraging timestamp or date information in the data, we can determine incremental changes since the last ingestion. For example, in a sales transactions table, extracting the maximum timestamp from previously ingested data allows us to retrieve only new or updated transactions that occurred after this timestamp. This approach minimizes computational resources and optimizes the ingestion process by processing only relevant data.

\textbf{Hash-Based Ingestion}: Utilizing hash functions to detect changes in the data ensures that only modified or new records are ingested. This technique is particularly useful when date columns are not available or not reliable for tracking changes.

\begin{figure} [!ht]
\centering
    \includegraphics[scale=0.35]{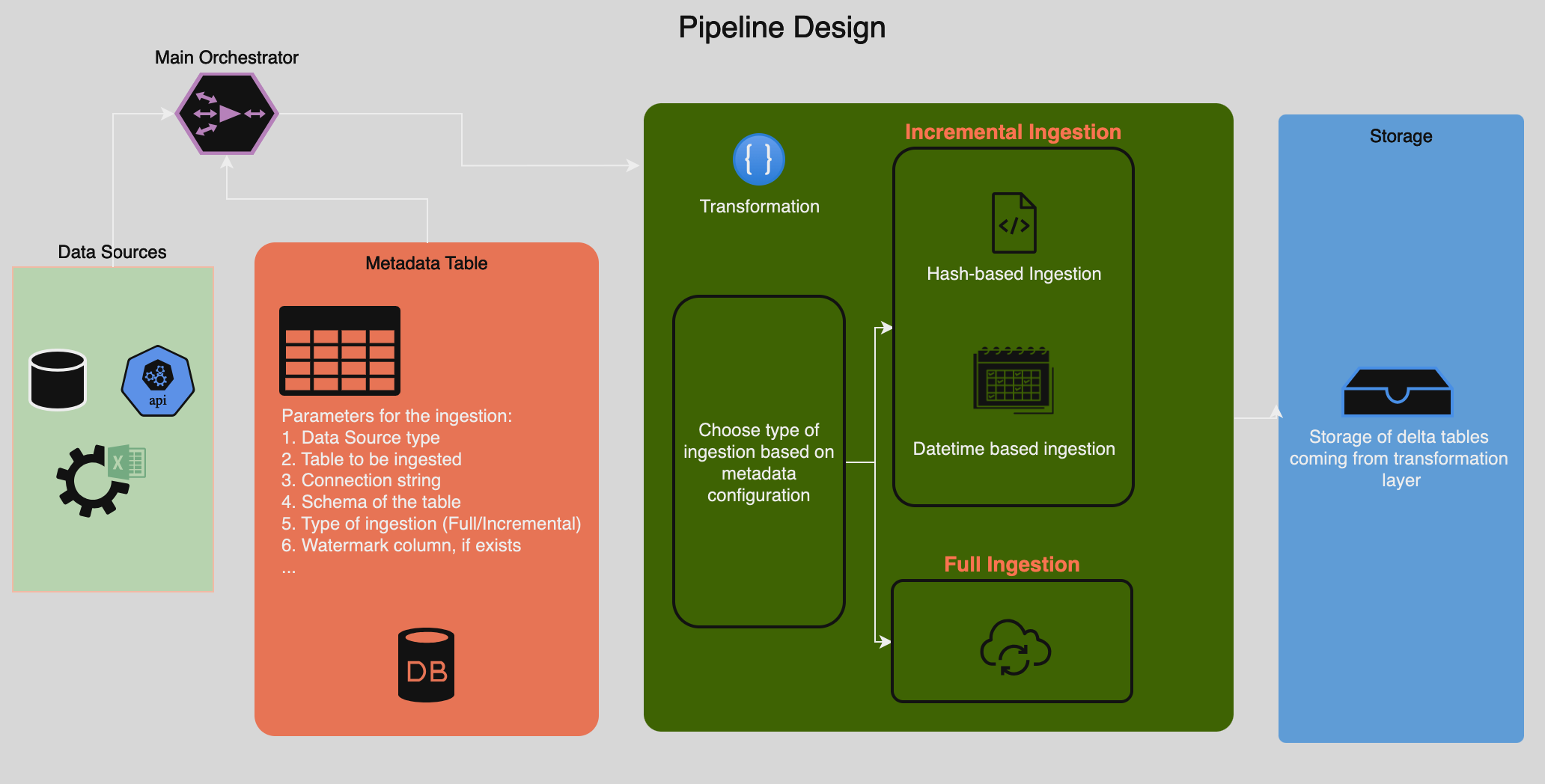}
    \caption{Example of the flow of the design pattern proposed. From the data source to the Storage}
    \label{fig:my_label4}
\end{figure}

In this proposed design pattern, the data ingestion pipeline is designed to be agnostic of specific cloud platforms, leveraging any suitable orchestrator such as Azure Data Factory, Google Cloud Composer, AWS Data Pipeline or any other. This flexibility allows for the construction of a unified pipeline that reads from the metadata table described earlier.
For each table to be ingested, the pipeline dynamically determines the ingestion technique—whether full refresh or incremental—based on the configuration specified in the metadata table. This approach ensures that the data ingestion process is adaptive and optimized for each data source's characteristics and update frequencies. Furthermore, the schema used for ingesting each table could also be governed by the metadata table. This parameterization ensures consistency and reliability across different ingestion scenarios. By centralizing these configurations in the metadata table, the pipeline becomes highly parameterized and easily adaptable to various use cases and environments.

The proposed design pattern is validated in our testbed using a specific cloud provider, demonstrating its feasibility, scalability, and flexibility.  It offers a versatile solution adaptable to different use cases, enhancing the efficiency and effectiveness of data ingestion processes in diverse organizational contexts.

\subsection{Testbed and validation}

To validate the effectiveness of our proposed design pattern, we utilized an Azure cloud infrastructure as our testbed. This infrastructure includes essential tools such as Azure Data Factory for orchestration, Azure Databricks with Apache Spark for data processing, a SQL database for metadata storage, and Azure Data Lake Gen2 for storing ingested data. This integrated setup allowed us to perform comprehensive tests across different scales of data, comprising datasets with 1K, 100K, and 1M rows respectively.

By leveraging Azure Data Factory, we seamlessly integrated the metadata-driven approach, where parameters such as data source, table name, ingestion method (incremental or full refresh), and other configurations were dynamically retrieved from the SQL database. This level of customization ensured that each data ingestion task could adapt to the specific characteristics and update frequencies of the data sources involved.

In terms of data storage and management, we adopted a folder-based organization within Azure Data Lake Gen2, with each table stored in its own folder. We utilized the Delta Lake format to capitalize on its transactional capabilities, ensuring efficient data storage, management, and retrieval.

Azure Databricks played a pivotal role in the actual data ingestion and processing phases. When initiating data ingestion process, our pipeline first verified the existence of tables based on the metadata configuration. If a table did not exist or required a full refresh, the pipeline overwrite or create the folder in the lake. Conversely, for tables set for incremental ingestion, only new or updated data since the last ingestion timestamp was processed using Spark's powerful filtering capabilities.

\begin{figure} [!ht]
\centering
    \includegraphics[scale=0.45]{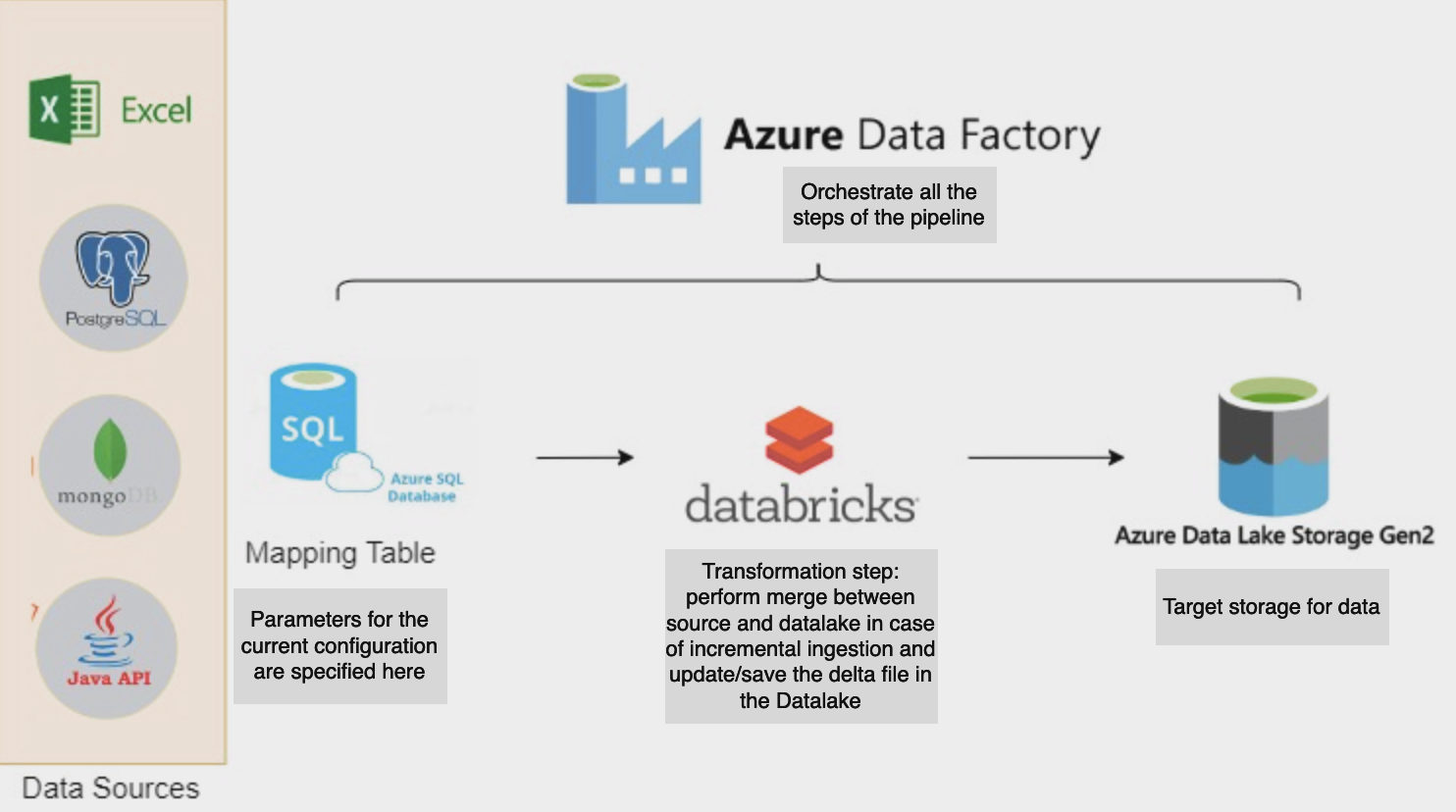}
    \caption{Schema of the Azure testbed used}
    \label{fig:my_label7}
\end{figure}

In our data ingestion process, we utilize two different techniques: ingestion based on a date column and ingestion based on a hash. These techniques, implemented using Azure Databricks and Apache Spark, provide flexibility and efficiency in handling data ingestion scenarios.

When ingesting data based on a date column, we leverage the timestamp or date information present in the data (and in the mapping table) to determine the incremental changes since the last ingestion. For example, let's consider a scenario where we have a table containing sales transactions. We can extract the maximum timestamp from the previously ingested data and use it as a reference point. Using Spark's filtering capabilities, we can retrieve only the new or updated transactions that have occurred after this timestamp. By reading and processing only the relevant data, we minimize the computational resources required and optimize the ingestion process.

Here an example of code to achieve this goal: 

\begin{lstlisting}[language=Scala, caption=Date-based Data Ingestion Script]
import org.apache.spark.sql.functions._

// Load previously ingested data
val previousData = spark.read.format("delta").load("/path/to/ingested_data")

// Find the maximum timestamp from the previous data
val maxTimestamp = previousData.select(max("timestamp"))
  .as[Long]
  .head()

// Read new data from the data source
val newData = spark.read.format("csv")
  .option("header", "true")
  .load("/path/to/new_data.csv")

// Filter new data based on the timestamp column
val filteredData = newData.filter(col("timestamp") > maxTimestamp)

// Perform transformations and process the filtered data
val processedData = filteredData.withColumn(...)
  // Apply necessary transformations

// Store the processed data into the Delta Lake
processedData.write.format("delta")
  .mode("append")
  .save("/path/to/ingested_data")
\end{lstlisting}

On the other hand, when employing ingestion based on a hash, we use a hash function to compare the incoming data with the existing dataset to identify changes. This technique is useful when a reliable timestamp column is not available or when the data source does not maintain timestamps consistently. For example, consider a customer profile table. We can generate a hash value for each record in the incoming dataset and compare it with the corresponding hash values in the existing table. If a record's hash value differs from the one in the table, it indicates a change or a new record. Using Spark's data comparison and transformation capabilities, we can identify and ingest only the modified or new records efficiently.

\begin{lstlisting}[language=Scala, caption=Hash-based Data Ingestion Script]
import org.apache.spark.sql.functions._

// Load previously ingested data
val previousData = spark.read.format("delta").load("/path/to/ingested_data")

// Read new data from the data source
val newData = spark.read.format("csv")
.option("header", "true")
.load("/path/to/new_data.csv")

// Generate hash values for the new data
val hashFunc = udf((col1: String, col2: Int) \
/* implement hash function here */)
val newDataWithHash = newData
.withColumn("hash", hashFunc(col("column1"), col("column2")))

// Compare the hash values to identify changes
val modifiedData = newDataWithHash
.join(previousData, Seq("id", "hash"), "left_anti")

// Perform transformations and process the modified data
val processedData = modifiedData.withColumn(...) 
// Apply necessary transformations

// Store the processed data into the Delta Lake
processedData.write.format("delta").mode("append")
.save("/path/to/ingested_data")
\end{lstlisting}

In the example code snippet provided, the \textit{hashFunc} at line 12, is a user-defined function (UDF) that implements the hash function. The function takes specific columns (\textbf{column1} and \textbf{column2} in the example) as input and computes the corresponding hash code. The resulting hash code is then compared to the hash codes of the previously ingested data to identify any differences or modifications.

By comparing the hash codes, you can efficiently determine which records in the new data have changed or are different from the previously ingested data. This enables you to selectively process and store only the modified or new records, avoiding redundant processing of unchanged data.

In both cases, once the incremental data is identified, we can use Spark's data manipulation functionalities to perform any necessary transformations, cleaning, or aggregations before storing the data in the Delta Lake format within the data lake. The Delta Lake format provides several benefits, such as transactional capabilities, schema enforcement, and efficient data processing.

To validate the effectiveness of the hybrid (incremental) ingestion method compared to full ingestion using the proposed design pattern, we conducted performance tests on a simulated dataset within our Azure-based testbed. To conduct tests on our data ingestion testbed, we used a synthetic dataset containing bank transactions. This dataset includes information such as transaction amount, origin, destination, and other transaction attributes. It consists of five tables of varying sizes: 1K rows, 10K rows, 100K rows, 1M rows and 1B rows, with each table featuring 25 columns of different data types, from integer, to varchar to bit type. Table \ref{tab:bank-transactions-condensed} shows Example Bank Transactions Dataset (Partial View). This setup allows us to simulate realistic data scenarios for evaluating data ingestion processes effectively.

\begin{table}[ht]
\centering
\renewcommand{\arraystretch}{1.5}
\begin{tabular}{|l|l|r|r|l|r|}
\hline
\textbf{Country} & \textbf{Sector} & \textbf{Gross Amount} & \textbf{Net Amount} & \textbf{Disbursed From} & \textbf{Disbursed To} \\
\hline
USA & Finance & 5000.00 & 4950.00 & Account A & Account B \\
\hline
UK & Retail & 120.50 & 100.00 & Account C & Merchant X \\
\hline
Germany & Economy & 10000.00 & 9900.00 & Account D & Account E \\
\hline
France & Technology & 800.00 & 750.00 & Account F & Supplier Y \\
\hline
Japan & Services & 2500.00 & 2450.00 & Account G & Customer Z \\
\hline
Brazil & Healthcare & 300.00 & 280.00 & Account H & Clinic W \\
\hline
\end{tabular}
\caption{Example Bank Transactions Dataset (Partial View)}
\label{tab:bank-transactions-condensed}
\end{table}

In this work, we're testing three ingestion methods on four identical tables of the same size, as specified. In the first case, the four tables are ingested entirely, representing full ingestion. In the second case, all tables are ingested using incremental methods. In the third case, two tables are ingested fully while the other two are ingested incrementally. This scenario reflects a data source that does not support incremental ingestion, possibly because it only provides one snapshot at a time without historical data. So the flexibility to configure it in the metadata table gives us the possibility to keep the incremental benefit for at least two tables.

We measured the execution times for full ingestion, incremental ingestion, and hybrid ingestion, implemented using hash comparison, assuming that at least 60\% of the table data remained unchanged between pipeline runs. The results are summarized in Table \ref{tab:execution-times-comparison} below:
\begin{table}[ht]
\centering
\renewcommand{\arraystretch}{1.5}
\begin{tabular}{|l|c|c|c|}
\hline
\textbf{Dataset Size} & \textbf{Incremental} & \textbf{Full} & \textbf{Hybrid} \\
\textbf{(\#rows)} & \textbf{Ingestion Time} & \textbf{Ingestion Time} & \textbf{Ingestion Time} \\
 & \textbf{(min)} & \textbf{(min)} & \textbf{(min)} \\ \hline
1K & 1.2 & 1 & 1 \\ \hline
10K & 2.4 & 3 & 2.7 \\ \hline
100K & 3.5 & 8 & 5 \\ \hline
1M & 4.7 & 14 & 7 \\ \hline
1B & 6 & 26 & 9.3 \\ \hline
\end{tabular}\caption{Comparison of Execution Times for Incremental, Full, and Hybrid Ingestion Methods}
\label{tab:execution-times-comparison}
\end{table}

\begin{itemize}
    \item 1K rows: Incremental ingestion takes approximately 1.2 minutes, slightly slower than full ingestion at 1 minute. Hybrid ingestion also takes 1 minute. The difference here is minimal due to the small dataset size, where all methods perform efficiently.
    \item 10K rows: Incremental ingestion continues to perform slightly better, taking 2.4 minutes compared to 3 minutes for full ingestion. Hybrid ingestion takes 2.7 minutes, reflecting the advantage of processing only the changed data in part of the dataset.
    \item 100K rows: Incremental ingestion takes 3.5 minutes, significantly quicker than full ingestion which requires 8 minutes. Hybrid ingestion takes 5 minutes, showing efficiency gains as it combines both ingestion methods.
    \item 1M rows: Incremental ingestion completes in 4.7 minutes, notably faster than full ingestion which takes 14 minutes. Hybrid ingestion takes 7 minutes, demonstrating substantial time savings due to targeted processing of incremental changes in part of the dataset.
    \item 1B rows: Incremental ingestion completes in 6 minutes, significantly faster than full ingestion at 26 minutes. Hybrid ingestion takes 9.3 minutes. Even with a massive dataset, hybrid ingestion maintains efficiency by combining full and incremental methods to process the data.
\end{itemize}

These results demonstrate that while both methods can handle smaller datasets with comparable efficiency, incremental ingestion becomes increasingly advantageous as dataset sizes grow. It minimizes unnecessary processing, reduces overall execution times, and optimizes resource utilization, making it a preferred choice for scalable data ingestion workflows. 

Our findings show that the hybrid mode is an efficient technique for data ingestion, particularly in scenarios where up-to-date information, system performance, and resource optimization are crucial. This method offers an optimal solution for data ingestion from various sources based on their specific requirements, whether full or incremental. The hybrid approach balances the fast execution and optimal use of resources on one hand, and the need for full data transfer on the other. Hybrid models leverage the strengths of both full and incremental ingestion, optimizing resource utilization and performance. This approach provides flexibility, allowing organizations to tailor their data ingestion strategy to their specific needs. 

Taking the scenario of ingesting a 1 billion rows, with some data sources requiring incremental mode and others requiring full mode, a purely full mode solution would take 26 minutes. Incremental mode alone cannot be used because some data sources require a full transfer of data. Therefore, it is evident that the hybrid mode is the best solution (9.3 minutes), meeting the requirements effectively while significantly reducing the time compared to the full mode. Additionally, a significant advantage of this approach is that data engineers do not need to build separate data pipelines for each data source. Instead, a single data pipeline can dynamically switch between modes based on a metadata table, thereby streamlining the ingestion process and reducing the complexity of pipeline management.

\section{Conclusions}

In conclusion, as demonstrated before, the combination of incremental and full refresh data ingestion techniques in a cloud-based architecture offers several benefits and provides a high level of customizability. 

Flexibility is the first critical benefit. Our metadata-driven approach allows for the dynamic adjustment of ingestion strategies based on the characteristics of each data source, providing the flexibility needed in diverse data ecosystems. 
Efficiency is enhanced through the combination of incremental and full refresh techniques. It's evident how efficient data ingestion techniques, such as incremental loading, reduce the computational burden and improve performance. 

The metadata-driven approach also provides high levels of customization and control. Sakr et al. \cite{sakr2011} advocate for metadata management to enhance data governance and control. By storing configuration parameters and ingestion types in a mapping table, organizations gain fine-grained control over their data ingestion processes, enhancing data quality and consistency as ingestion methods can be tailored to the specific needs of each data source. 

A cloud-agnostic capability is another significant advantage. A report by Gartner (2020) \cite{gartner2020} on cloud data management emphasizes the need for cloud-agnostic solutions to avoid vendor lock-in and ensure interoperability across different platforms. The proposed design pattern’s principles enable deployment across various cloud environments, protecting the organization from being tied to a single cloud provider and facilitating multi-cloud strategies and migrations if necessary.
Lastly, enhanced data management practices are integral to the design pattern. The folder-based approach and use of transactional data formats like Delta Lake ensure data is well-organized and easily retrievable. This improves data accessibility and reliability, supporting analytics and decision-making processes.

By integrating these benefits from literature, the proposed design pattern not only addresses the technical challenges of data ingestion but also aligns with best practices and advances in data management technologies. This makes it a valuable contribution to the literature and applicable to a wide range of use cases, offering a versatile solution adaptable to different organizational contexts.

Overall, the combination of incremental and full refresh ingestion techniques, along with the use of customizable mapping tables and Azure services, empowers us to optimize our data processing pipeline. We can adapt to different data source capabilities, minimize processing time and resource usage, ensure data consistency and integrity, and ultimately enhance the overall efficiency and effectiveness of our data ingestion process in the cloud-based architecture.

In conclusion, this paper provides a comprehensive framework for designing and implementing a robust and flexible data ingestion pipeline in a cloud-based data lakehouse architecture. By following the outlined approach and utilizing the recommended tools and techniques, organizations can efficiently ingest, process, and store data from various sources, paving the way for data-driven success and innovation.

\printbibliography

\end{document}